\documentclass[12pt]{article}
\usepackage{amsfonts,amssymb}
\begin{document}
\title {Five-dimensional general relativity and Kaluza-Klein theory}
\author{Valentin D. Gladush\\
Department of Theoretical Physics, Dnepropetrovsk National University,\\
per. Nauchniy 13, Dnepropetrovsk 49050, Ukraine}
\date{\today}

\maketitle
\begin{abstract}
We consider 5D spaces which admit the most symmetric 3D subspaces. 5D
vacuum Einstein equations are constructed and 5D analog of the mass
function is found. The corresponding conservation law leads to 5D analog
of Birkhoff's theorem. Hence the cylinder condition is dynamically
implemented for the considered spaces. For some obtained metrics a period
of space with respect to the fifth coordinate was found. The problem of
the dynamical degrees of freedom of the fields system obtained in the
process of dimensional reduction is discussed, and the problem of their
interpretation is considered. One can think that the parametrization of
the scalar field and 4D metric leading to the conformally invariant 4D
theory for interacting gravitational and scalar fields is most natural and
adequate.
\end{abstract}

\section{Introduction}
The basic proposal of Kaluza-Klein theory is the cylinder condition.
However it is one of those assumptions which generates many questions and
requires some justification. Other proposal is the idea of compactifying
the fifth dimension. There are also some open questions here, one of which
is the magnitude of the period of 5D space $V^5$ with respect to fifth
coordinate.

On the other hand in General Relativity the following result is well-known
(Birkhoff's theorem): under certain conditions, the Schwarzschild solution
is a unique spherically symmetric solution of vacuum Einstein equations
\cite {Kram1}. This theorem is in fact the consequence of existence of the
mass function and the corresponding conservation law. In 5-dimensional
(5D) General Relativity an alleviated version of Birkhoff's theorem
\cite{Bron} is valid: the static, spherically symmetric solution of 5D
vacuum Einstein equations is unique \cite {Kram2}, but there is a variety
of non-stationary vacuum solutions \cite {Krech}.

It turns out that the above mentioned questions are connected. Expanding
the concept of the spherical symmetry we can introduce 5D analog of the
mass function and formulate 5D analog of Birkhoff's theorem. For this
purpose it is enough to perform replacement: $\{V^4,\,O(2),\,S^2\}
\rightarrow \{V^5,\,O(3),\,S^3\}$ in the definition of spherical symmetry,
where $O(2)$ and $O(3)$ are the rotation groups. They are the groups of
motion $V^4 $ and $V^5 $ which are transitive on 2D and 3D spheres $S^2 $
and $S^3 $ respectively. In this meaning we can pose the problem of
existence and uniqueness of 5D analog of the Schwarzschild solution. Hence
we come to the cylinder condition and to the period of 5D space $V^5$ with
respect to fifth coordinate.

In this paper we consider the more general case, when 5D space $V^5 $
admits a transitive action of the isometry group on 3D spacelike or
timelike surfaces $\Sigma^3$ of constant curvature. In the framework of 5D
General Relativity we construct 5D vacuum Einstein equations, and find 5D
analog of the mass function. Hence we obtain the conservation law which
points to the fact that 5D analog of the Birkhoff's theorem takes place.
Thus we come to the conclusion that in the generalized curvature
coordinates all quantities do not depend on the fifth coordinate. The last
means that for such spaces the cylinder condition is implemented
dynamically. For some obtained metrics we find the period of space with
respect to fifth coordinate. The problem of the dynamical degrees of
freedom for the system of the interacting scalar and gravitational fields
which obtained in the process of dimensional reduction is discussed. The
separation problem of dynamical degrees of freedom and their
interpretations is considered. The various representations of the obtained
metrics are discussed.

\section{Birkhoff's theorem in five-dimensional gravity}
In the framework of 5D General Relativity \cite {wes} let us consider a
pseudo-Riemannian space $V^5$ with the metric $^{(5)}g_{AB}$ ($A,~B=0$,\
1,\ 2,\ 3,\ 4) (signature $(+\ -\ -\ -\ -)$) which generally depends on
all coordinates $x^A $. The metric satisfies 5D vacuum Einstein equations
\begin{equation}\label{2.1}
  ~^{(5)}G^{A}_{B}=\,^{(5)}\!R^{A}_{B}
  -\frac{1}{2}\ \,^{(5)} \!R\ \delta^{A}_{B}=0 .
\end{equation}
These equations can be derived by varying 5D version of the usual
Einstein-Hilbert action:
\begin{equation}\label{act}
     I=-\int \sqrt{|^{(5)}g|}~^{(5)}R\ d^5 x \,,
\end{equation}
where $^{(5)}g = \det||^{(5)}g_{AB}||$.

We consider the spaces $V^5$ which are the generalization of the
spherically-symmetric spaces of the General Relativity. These spaces admit
the maximally symmetric 3D surfaces $\Sigma^3$. For generality, we shall
consider both spatial and time-spatially surfaces $\Sigma^3$. Thus $V^5 $
admits a transitive action of the isometry group on 3D spacelike or
timelike surfaces $\Sigma^3$ of constant curvature.

The desired metric can be represented in the following 2+3 form
\begin{equation}\label{2.2}
^{(5)}ds^2 =^{(5)}g_{AB}dx^A dx^B =g_{ab}dx^a dx^b -
{\Lambda^2}\,^{(3)}d\Omega^2\,.
\end{equation}
where, according to \cite{ei} (for more details see Appendix A),
\begin{equation}\label{2.3b}
   ^{(3)}d\Omega^2 = h_{ij}dx^i dx^j
   = \frac{\epsilon(dx^1)^2 + (dx^2)^2 + (dx^3)^2}
   {\left(1+\frac{K_{0}}{4}\,S^2 \right)^2}
\end{equation}
is the metric of 3D space of the unit positive $(K_{0}=1)$, negative
$(K_{0}=-1)$ or zero $(K_{0}=0)$ curvature. Here $S^2 =\epsilon(x^1)^2
+(x^2)^2 +(x^3)^2$, $\epsilon =1$ for the spatial section $\Sigma^3$ and
$\epsilon =-1$ for the time-spatially section $\Sigma^3 $. The quantities
$g_{ab}$ and $\Lambda$ depend on the coordinates $x^a$ ($a, b=0, 4$; $\ i,
j=1, 2, 3$) only.

The components of the Ricci tensor for the metric (\ref{2.2}) have the
form:
\begin{eqnarray}
    ~^{(5)}R_{ab} & = & ^{(2)}R_{ab}
    -3\Lambda^{-1}\nabla_{a}\nabla_{b}\Lambda\, ,
    \label{rab}\\
    ~^{(5)}R_{ai} & \equiv & 0\, ,    \label{rai}\\
    ~^{(5)}R_{ik} & = & [\Lambda\Delta\Lambda +2(\nabla\Lambda)^2
    +2K_{0}]h_{ik}\, ,    \label{rik}
\end{eqnarray}
where $~\Delta =\nabla^{a}\nabla_{a}, \ (\nabla\Lambda)^2 =
g^{ab}\nabla_{a}\Lambda\nabla_{b}\Lambda$,  $\nabla_{a}$ is the covariant
derivative with respect to coordinate $x^a$ calculated with the help of
the 2D metric $g_{ab}$, $~^{(2)}R_{ab}$ is the Ricci tensor and $~^{(2)}R$
is the curvature scalar of 2D space with the metric $g_{ab}$. Hence,
according to (\ref {2.1}) we obtain the equations of 5D gravitation for
the metric (\ref{2.2}):
\begin{eqnarray}
    ~^{(5)}G^{a}_{b} & = & -\frac{3}{\Lambda}\nabla^{a}\nabla_{b}\Lambda
    +\frac{3}{\Lambda^2}\left(\Lambda\Delta\Lambda +(\nabla\Lambda)^2
    +K_{0}\right)\delta^{a}_{b}=0\, ,     \label{2.5}\\
    ~^{(5)}G^{i}_{k} & = &\left(\frac{1}{2}\,\Lambda^2~^{(2)}R
    -2\Lambda\Delta\Lambda - (\nabla\Lambda)^2
    - K_{0}\right)\delta^{i}_{k}=0\, .   \label{2.7}
\end{eqnarray}
From these equations it follows that
\begin{equation}\label{2.8}
   2\Lambda^3\left(^{(5)}G^{\,a}_{a}\Lambda_{,b}
   -~^{(5)}G^{a}_{b}\Lambda_{,a}\right) =
   3\left(\Lambda^2(\nabla\Lambda)^2
   +K_{0}\Lambda^2\right)_{,\,b} = 0\, .
\end{equation}
Hence one finds the conservation law
\begin{equation}\label{2.9}
 G \equiv \Lambda^2 (\nabla\Lambda)^2 +K_{0} \Lambda^2 =
 \mbox{const}\, ,
\end{equation}
which points to the fact that 5D analog of the Birkhoff's theorem takes
place. 4D analog of (\ref{2.9}) corresponds to the conservation law of the
complete mass inside a collapsing ball and determines the mass function
\cite{her}. In 5D case it corresponds to the conservation of quantity
being the integral characteristic of the sources of some scalar field.
Therefore, the quantity $G$ can be named as charging function and
corresponds to the conservation law of the scalar charge in 5D General
Relativity.

From the equations (\ref{2.1}), (\ref{rik}) it follows that
  $\Lambda\Delta\Lambda +2(\nabla\Lambda)^2 +2K_{0} =0 $.
Hence, using (\ref{2.9}) one obtains
\begin{equation}\label{Delt}
  \Delta \Lambda + \frac{2G}{\Lambda^3}=0\, .
\end{equation}
Let us consider now the regions in which
$\mbox{sign}(\nabla\Lambda)^2=\mbox{sign}\,\epsilon$. Using the admissible
transformations $x^a=x^a(\tilde{x}^b)$ one can choose the coordinates
$\tilde{x}^b$ so that $\widetilde{g}_{04}=0$ and $\widetilde{x}^0=\Lambda
$. The new coordinates $\{\widetilde{x}^0=\Lambda,~\widetilde{x}^4\equiv
Z\}$ are the analog of the curvature coordinates in General Relativity. As
a result 5D interval (\ref{2.2}) can be rewritten in the form
\begin{equation}\label{2.10}
   ~^{(5)}ds^2 = -M^2 dZ^2 + \epsilon N^2 d\Lambda^2
    - {\Lambda^2}\,^{(3)}d\Omega^2\, .
\end{equation}
Then from the conservation law (\ref{2.9}) it follows that
\begin{equation}\label{2.11}
  N^{-2}=-\, \epsilon K_{0} +\frac{\epsilon\,G}{T^2}\, .
\end{equation}
With the help of (\ref{Delt}) we find $\partial (M N)/\partial \Lambda
=0$. Hence $M=N^{-1}f(Z)$, where $f(Z)$ is an arbitrary function. One can
suppose, without loss of generality, that $f(Z)=1$. As a result we obtain
the metric
\begin{equation}\label{2.14a}
   ~^{(5)}ds^2 = -\left(-\epsilon K_{0}
   +\frac{\epsilon G}{\Lambda^2}\right)dZ^2
   + \frac{\epsilon d\Lambda^2}{-\epsilon K_{0}
   +\frac{\epsilon G}{\Lambda^2}}
   - \Lambda^2\,\frac{\epsilon (dx^1)^2 +(dx^2)^2 +(dx^3)^2}
   {\left(1+\frac{K_{0}}{4}\,S^2 \right)^{2}}\, ,
\end{equation}
which essentially proves 5D Birkhoff's theorem for $V^5$.

It is easy to see that the obtained metric has the singular null
hypersurfaces $\Lambda=\pm\sqrt{K_{0} G}$ when $K_{0}G>0$. In some sense
they are similar to the Schwarzschild event horizon. When $\epsilon =-1$
the regularity condition of the sections $x^i=\mbox{const}$ on these
horizons leads to the period $L=2\pi\sqrt{K_{0} G}$ of $V^5$ with respect
to fifth coordinate. In the case $\epsilon =1$, considering the
prolongation of the metric on the imaginary axis $Z=\imath\, Z'$, we
obtain the same period for $Z'$.

In the case of $\epsilon =1$ setting $\Lambda=T $, we have
\begin{equation}\label{2.14T}
   ~^{(5)}ds^2 = -\left(-K_{0} +\frac{G}{T^2}\right)dZ^2
   + \frac{dT^2}{-K_{0} +\frac{G}{T^2}}
   - T^2\,\frac{(dx^1)^2 +(dx^2)^2 +(dx^3)^2}
   {\left(1+\frac{K_{0}}{4}\,S_{+}^2 \right)^{2}}\, .
\end{equation}
This metric is 5D analog of the Schwarzschild solution. It admits the
rotation group $O(3)$. We can fix the signs $K_{0}$ and $G$  by requiring
the space $V^5$ to admit 4D flat sections. As follows from Sec.3 it is
possible when $K_{0}=-1$ and $G<0$.

For the sections $\Sigma^3$ of positive curvature $(K_{0}=1)$ the
condition $g_{TT}>0$ is possible when $G>0$ and it leads to the finiteness
of the model in time $-\sqrt{G}<T< \sqrt{G}\,$. For the sections
$\Sigma^3$ of negative curvature $(K_{0}=-1)$ the metric (\ref{2.14T})
will be regular for all $T\neq 0$ if the condition  $G>0 $ is satisfied.
If $G<0$, the inequality $g_{TT}>0$ will lead to two domains $T<
-\sqrt{-G}$ and $ T>\sqrt{-G}$.

The other solutions can be obtained by an analytic continuation the
solution (\ref{2.14T})  through the null hypersurfaces
$T=\pm\sqrt{K_{0}G}$. In this case the sense of coordinates $Z$ and $T$
varies. Therefore it is necessary to perform the replacement $
Z\rightarrow T, \,T\rightarrow Z $. As a result the dependence of metric
on $Z$ appears. This situation is similar to transition through the
Schwarzschild horizon. Here we do not consider these regions. The other
variant of the metric is possible under replacement $Z\rightarrow
T,\,T\rightarrow R,\,x^1\rightarrow x^4=z$. Then we come to the metric
\begin{equation}\label{tan}
   ~^{(5)}ds^2 = \left(K_{0} -\frac{G}{R^2}\right)dT^2
   -\frac{dR^2}{K_{0} -\frac{G}{R^2}}
   - R^2\,\frac{(d z)^2 +(dx^2)^2 +(dx^3)^2}
   {\left(1+\frac{K_{0}}{4}\,S_{+}^2 \right)^{2}}\, .
\end{equation}
It can be transformed to the Tangerlini metric \cite{tang}. Here the fifth
coordinate and two spatial coordinates are associated by symmetry $O(3)$.
Therefore, when a rotation from this group takes place, they are
transforming jointly. However, by the data, the fifth coordinate should be
transformed in each point $V^4 $ independently on the space-time
coordinates as a coordinate of the internal symmetry space. That is why
this solution is inconsistent with statement of the problem and we do not
also consider it here.

In the case of $\epsilon =-1$ we suppose $\Lambda=R$, $x^1\rightarrow x^0=
t$ and the metric $V^5$ acquires the form
\begin{equation}\label{2.14b}
   ~^{(5)}ds^2 = -\left(K_{0} -\frac{G}{R^2}\right)dZ^2
   - \frac{dR^2}{K_{0} -\frac{G}{R^2}}
   + R^2\,\frac{(dt)^2 -(dx^2)^2 -(dx^3)^2}
   {\left(1+\frac{K_{0}}{4}\,S_{-}^2 \right)^{2}}\, .
\end{equation}
For the sections $\Sigma^3$ of positive curvature $(K_{0}=1)$ when $G>0$
the condition $g_{RR}<0$ performs for the exterior regions $R<-\sqrt{G}$
and $ R>\sqrt{G}$ of the space $V^5$. In case that $G<0$ the inequality
$g_{RR}<0$ is possible for all $R$. For the sections $\Sigma^3 $ of
negative curvature $(K_{0}=-1)$ the condition $g_{RR}<0$ is possible for
the interior region $-\sqrt{-G}<R< \sqrt{-G}$ when $G<0$. There are the
singular null hypersurfaces $R=\pm R_{G}=\pm \sqrt{K_{0}G}$ when
$K_{0}G>0$, however transition through these hypersurfaces leads to the
non-physical metric.

By analogy with $t$-independent (static character) Schwarzschild solution
in the $R$-domain, the metrics (\ref{2.14T}), (\ref{2.14b}) are
independent on the fifth coordinate $x^4=Z$. Besides, provided $K_{0}G>0$,
this coordinate (or its imaginary prolongation for the metric
(\ref{2.14T})) has the period $L=2\pi\sqrt{K_{0}G}$ for the metric
(\ref{2.14b}). Thus Kaluza's cylinder condition is implemented dynamically
here. It explains also why the fifth dimensionality is compact and gives
the value of its period.

\section{Some representations of obtained solutions}

It turns out that the metric (\ref{2.14a}) admits conformally flat 4D
sections. Indeed, after substitution
\begin{equation}\label{v3.7}
  \Lambda=U \left(1 - \frac{\epsilon G}{4U^2}\right)\, ,
\end{equation}
under condition $K_{0}\epsilon =-1$, the metric is converted as:
\begin{eqnarray}\label{v3.8}
    ~^{(5)}ds^{2} =
       - \left(\frac{1+\frac{\epsilon G}{4U^2}}
       {1-\frac{\epsilon G}{4U^2}}\right)^2 dZ^2
       + \left(1-\frac{\epsilon G}{4U^2}\right)^2 \left(\epsilon dU^2
       - U^2\,^{(3)}d\Omega^2\right)\, .
\end{eqnarray}
In the case $\epsilon G<0$ formula (\ref{v3.7}) defines the maps of the
exterior region $\sqrt{-\epsilon G}<\Lambda<\infty$ 3-spheres
$\Lambda=\sqrt{-\epsilon G}$ onto the region $-\infty<U<\infty$. The
metric (\ref{v3.8}) is invariant with respect to the reflection
$U\rightarrow -U$ and the inversion $U\rightarrow U'=\epsilon G/4U$ of the
coordinate $U$.

With the help of the formulas (\ref{a2met}), (\ref{a2split}), where we
suppose $\epsilon_{\mu \nu}=-\eta_{\mu\nu}$ $(\mu, \nu = 0,\,1,\,2,\,3)$,\
$e\rightarrow \epsilon $,\ $\Lambda\rightarrow U$, the interval
(\ref{v3.8}) can be rewritten as
\begin{eqnarray}\label{v3.9}
    ~^{(5)}ds^{2} =
       - \left(\frac{1+\frac{\epsilon G}{4U^2}}
       {1-\frac{\epsilon G}{4U^2}}\right)^2 dZ^2
       + \left(1-\frac{\epsilon G}{4U^2}\right)^2
       \eta_{\mu\nu}dy^{\mu}dy^{\nu}\, ,
\end{eqnarray}
where $U^2=\eta_{\mu\nu}y^{\mu}y^{\nu}$, and $\eta_{\mu\nu}$ is the
Minkowski metric. The metric (\ref{v3.9}) is invariant with respect to
conformal transformations which in addition to the Lorentz transformation
contain inversion operation: $$ y^{\mu}=\frac{\epsilon
G}{4}\frac{y'^{\mu}}{U'^2}\,, \qquad  (U'^2
=\eta_{\alpha\beta}y'^{\alpha}y'^{\beta})\,.$$

When $K_{0}\epsilon =1$, from the condition $g_{ZZ}<0$, it follows that
$\epsilon G>0$. Then after the substitutions
\begin{equation}\label{tran1}
  \Lambda = \sqrt{\epsilon G}\,\cos\ln\left(\frac{U}{\sqrt{\epsilon G}}\right)\,, \quad
  \Lambda = \sqrt{\epsilon G}\,\sin\ln\left(\frac{U}{\sqrt{\epsilon G}}\right)
\end{equation}
the metric (\ref{2.14a}) can be written in the forms
\begin{eqnarray}
    ~^{(5)}ds^{2} =
       - \tan^{2}\ln\left(\frac{U}{\sqrt{\epsilon G}}\right)\, dZ^2
     +\frac{\epsilon G}{U^2}\cos^2\ln\left(\frac{U}{\sqrt{\epsilon G}}\right)
      \left(\epsilon dU^2 - U^2\,^{(3)}d\Omega^2\right)\,, \label{v3.8a}\\
       ~^{(5)}ds^{2} =
       - \cot^{2}\ln\left(\frac{U}{\sqrt{\epsilon G}}\right)\, dZ^2
     +\frac{\epsilon G}{U^2}\sin^2\ln\left(\frac{U}{\sqrt{\epsilon G}}\right)
      \left(\epsilon dU^2 - U^2\,^{(3)}d\Omega^2\right)\,. \label{v3.8b}
\end{eqnarray}
respectively.

For the case $G<0$ and $K_{0}\epsilon =-1$ the metric (\ref{2.14a}) admits
also flat 4D space-time slices which are non-orthogonal to the coordinate
lines $x^5=Z$. Indeed, it may be rewritten in the form similar to
Painlev\'{e} representation \cite{pain} of the Schwarzschild solutions.
This can be performed with the help of transformation of the fifth
coordinate
\begin{equation}\label{trans5}
  Z = z + \int\left(\frac{\Lambda}{\Lambda_G}-
  \epsilon \frac{\Lambda_G}{\Lambda}\right)^{-1}d \Lambda\,,
\end{equation}
where  $\Lambda_{G}=\sqrt{-G}$. As a result the metric (\ref{2.14a}) can
be rewritten as
\begin{equation}\label{penl}
   ~^{(5)}ds^2 = -dz^2
   + \epsilon \left(d \Lambda-\epsilon \frac{\Lambda_G}{\Lambda}\,d z \right)^2
  - {\Lambda^2}\,^{(3)}d\Omega^2\, .
\end{equation}
At last, with help of the formulas (\ref{a2met}), (\ref{a2split}) of
appendix where $\epsilon_{\mu \nu}=-\eta_{\mu\nu}$,\ $e=\epsilon $, the
interval (\ref{penl}) can be written in the form
\begin{equation}\label{penl2}
   ~^{(5)}ds^2 = -dz^2 + \eta_{\mu\nu}
   \left(dy^{\mu}- \epsilon \frac{\Lambda_G}{\Lambda}\,\eta^{\mu}d z\right)
   \left(dy^{\nu}- \epsilon \frac{\Lambda_G}{\Lambda}\,\eta^{\nu}d z\right)\,.
\end{equation}
Here $\Lambda^2 =\epsilon \eta_{\mu\nu}y^{\mu}y^{\nu}$,
$\eta^{\mu}=y^{\nu}/\Lambda$,\  $d\Lambda =\epsilon \eta_{\mu}dy^{\mu} $.
Hence it can be seen that in the coordinates under consideration 4D
physical space is the set of flat space-time sections of a normal geodesic
congruence of the curves in $V^5 $ with a field of tangential vectors
$U^{A}=\{U^5 =1,\ U^{\mu}=\epsilon (\Lambda_G /\Lambda)\eta^{\mu}\}$. It
also follows from (\ref{penl2}) that the singularity $\Lambda =\Lambda_G $
in the metric (\ref{2.14a}) is stipulated by a choice of the coordinates
and is associated with incompleteness of the curvature coordinates
$\{Z,\Lambda \}$ similarly to the singularity of the event horizon $R=R_g
$ in Schwarzschild metric. However, in contrast to the curvature
singularity $R=0$ of the Schwarzschild solution, the curvature singularity
$\Lambda=0$ is situated on the light cone $(x^0)^2-(x^1)^2-(x^2)^2-(x^3)^2
=0$.

Note that in case of the sections $\Sigma^3 $ of zero curvature
$(K_{0}=0)$ the metric (\ref{2.14T}) has other simple representations.
Taking into account the condition $g_{ZZ}<0$, we suppose $G>0$. Then,
after replacement $T=x^0\sqrt{G}$, this metric becomes
\begin{equation}\label{3.1}
   ~^{(5)}ds^2 =-(x^0)^{-2}dZ^2 +G (x^0)^2 \left((dx^0)^2 -
  (dx^1)^2 +(dx^2)^2 +(dx^3)^2\right)\, .
\end{equation}
In case of the metric (\ref{2.14b}) for $K_{0}=0$ it is necessary to put
$G<0$. Then after replacement $G=-\widetilde{G},\ R\rightarrow x^1
\sqrt{\widetilde{G}},\ x^1 \rightarrow x^0 $ we have
\begin{equation}\label{3.1a}
   ~^{(5)}ds^2 =-(x^1)^{-2}dZ^2 +\widetilde{G} (x^1)^2 \left((dx^0)^2 -
  (dx^1)^2 +(dx^2)^2 +(dx^3)^2\right)\,.
\end{equation}
These metrics are 5D analogs of Kasner metric (the degenerate case). After
the change $x^0=2\sqrt{\tau\tau_0}$, $Z=2\tau_0 y$ and $G=(2\tau_0)^{-2}$
expression (\ref{3.1}) coincide with metric of the cosmological model
considered in \cite{chod}.

\section{On the separation of dynamical degrees of freedom}

In order to separate the space-time dynamical variables from dynamical
variable of the interior space in some solutions of 5D gravitation one
must construct (4+1)-split of 5D metric and perform the relevant
transformation of 4D metric and scalar field resulting from dimensional
reduction. There are no special problems at the first step, since the
split methods are well known (for example, see \cite{gl1} and references
therein) and metric is sufficiently ordinary. Further, a problem of a
conformal rescaling or conformal gauge is emerged. It is associated with
conformal ambiguity of the physical metric and scalar field on $V^4 $ and
with a possibility of their conformal transformation.

In case of $V^5 $ under consideration the dynamical system is in the state
with one conserved nonzero quantity only. In configurational space it
corresponds to one dynamical degree of freedom which can be connected
either with gravitational field or with some scalar field. Therefore we
shall consider the basic metric ansatzs of the Kaluza-Klein theory from
this point of view.

Let us rewrite the interval (\ref{2.2}) in (4+1)-form
\begin{equation}\label{met}
^{(5)}ds^2 = g_{\mu\nu}dx^\mu dx^\nu - W^2 dz^2\,,
\end{equation}
As it was shown above, for the considered symmetric case the cylinder
condition is implemented dynamically. Therefore quantities $W$ and
$g_{\mu\nu}$ does not depend on the coordinate $z$, and hence for 5D
action (\ref{act}) we have
\begin{equation}\label{act1}
     I=-L\int \sqrt{-^{(4)}g}\ W^{(4)}R\ d^4 x \,,
\end{equation}
where $L$ is the period of $V^5$ with respect to fifth coordinate,
$^{(4)}R$ is the scalar curvature of 4D space with metric $g_{\mu\nu}$.

We are free to perform the transformation $$W = W(\varphi), \quad
g_{\mu\nu}= f(\varphi)\widetilde{g}_{\mu\nu}\,. $$ Therefore the metric of
the physical space-time is determined up to a conformal factor. The
transformation should be chosen so that it was possible ``to
orthogonalize'' in some sense the action (\ref{act1}) and to separate the
degrees of freedom corresponding to the scalar and gravitational field.
One of the most popular ansatzs (see, for example, \cite{duff} and
\cite{gl2}) is
\begin{equation}\label{4.1}
^{(5)}ds^2 =e^{\varphi/\sqrt{3}} \widetilde{g}_{\mu\nu}dx^\mu dx^\nu -
e^{-2\varphi/\sqrt{3}}\,dz^2\, .
\end{equation}
In terms of new variables the action (\ref{act1}) reduces to the action
for the gravitational field $\widetilde{g}_{\mu\nu}$ and the scalar field
$\varphi$ with minimal coupling
\begin{equation}\label{act2}
     I=-L \int d^4 x \sqrt{-^{(4)}g} \left\{^{(4)} \widetilde{R}
     - \frac{1}{2} \widetilde{g}^{\mu\nu}\varphi_{,\mu}
     \varphi_{,\nu}\right\} \,.
\end{equation}
Comparing together the metrics (\ref{4.1}) and (\ref{2.14T}) we obtain the
scalar field and the new metric
\begin{equation}\label{phy}
    \varphi=-\frac{\sqrt{3}}{2}\ln\left(- K_{0}+\frac{G}{T^2}\right)\,, \quad
     \widetilde{g}_{\mu\nu}= g_{\mu\nu}\sqrt{- K_{0}+\frac{G}{T^2}}\,,
\end{equation}
where the metric $g_{\mu\nu}$ can be obtained from the solution
(\ref{2.14T}). The new fields $\varphi$ and $\widetilde{g}_{\mu\nu}$ are
functionally dependent and therefore can not represent the independent
dynamical degrees of freedom of the system. On that ground we should
reject the ansatz (\ref{4.1}) as far as it does not correspond to the
optimal representation of the dynamical system with one degree of freedom.

Now we consider the other ansatz \cite{gl2}
\begin{equation}\label{int}
       ~^{(5)}ds^{2} =
       \left(1+\frac{\psi}{\sqrt{6}}\right)^2
       ~^{(4)}ds'\,^{2}
       - \left(\frac{1-\frac{\psi}{\sqrt{6}}}
       {1+\frac{\psi}{\sqrt{6}}} \right)^2 dz^2  \, .
\end{equation}
For this case the expression (\ref{act1}) leads to the action
\begin{equation}\label{4.4}
  ~^{(4)}I = -L \int d^4 x \sqrt{-^{(4)}g} \left\{\left(1-\frac{\psi^2}
  {6}\right)R'- g'\,^{\mu\nu}\psi_{,\mu}
     \psi_{,\nu} \right\}\,,
\end{equation}
which describes the interacting gravitational $g'_{\mu\nu}$ and
conformally invariant scalar $\psi$ fields. The equations of motion for
new system have the form
\begin{eqnarray}
 & (\Delta - \frac{1}{6}R') \psi = 0\, ,
  \label{4.6}\\
 & G'_{\mu\nu} = 4\pi t_{\mu\nu}\equiv
  4\pi T_{\mu\nu}+  \frac{1}{6}\left(G'_{\mu\nu}
   -\nabla_{\mu}\nabla_{\nu}
   +g'_{\mu\nu}\Delta\right)\psi^{2}\, , \label{4.8}\\
 &4\pi T_{\mu\nu}=\psi_{\mu}\psi_{\nu}
  -\frac{1}{2}\,g'_{\mu\nu}(\nabla \psi)^2 \, , \label{4.9}
\end{eqnarray}
where $t_{\mu\nu}$ is the conformally invariant energy-momentum tensor of
scalar field $\psi$ with conformal coupling,  $G'_{\mu\nu}=
R'_{\mu\nu}-\frac{1}{2}\,g'_{\mu\nu}\,R'$, $R'_{\mu\nu}$ and $R'$ are the
Ricci tensor and the curvature scalar respectively,
$\Delta=\nabla^{\mu}\nabla_{\mu}$, $\nabla_{\mu}$ is the covariant
derivative with respect to the coordinate $x^{\mu}$. Here all the
quantities are calculated with the help of 4D metric $g'_{\mu\nu}$.

The metrics (\ref{int}) and (\ref{v3.9}) have the same form. Hence, as a
result of the comparison, we obtain the scalar field and the physical
metric
\begin{equation}\label{phy1}
    \psi=-\frac{\sqrt{6}}{4}\,\frac{\epsilon\, G}
    {(x^0)^2-(x^1)^2-(x^2)^2-(x^3)^2}\,,
    \quad g'_{\mu\nu}=\eta_{\mu\nu}\,.
\end{equation}
It is interesting that the conformal energy-momentum tensor vanishes
$t_{\mu\nu}=0$ for the obtained solution of the equations
(\ref{4.6}-\ref{4.9}). The received representation connects the single
degree of freedom of the system (which associated with conserved charge
$G$) with the conformally invariant scalar field $\psi$. It corresponds to
the condition of a separating problem of degrees of freedom. Here the
physical space-time is flat, and the scalar field $\psi$ is the classical
ghost. It has zero energy density and does not render influence on the
space-time.

It is easy to generalize the considered ansatz to a case of the
electromagnetic field $A_{\mu}$ \cite{gl2} when instead of (\ref{met}) we
use the metric
\begin{equation}\label{meta}
^{(5)}ds^2 = g_{\mu\nu}dx^\mu dx^\nu - W^2 (d z+A_{\mu}dx^{\mu})^2\,.
\end{equation}
Gauge transformations $  z=z' + f(x^{\mu}), \ A_{\mu}=A'_{\mu}-f_{,\mu}$
leave this metric invariant.

Now let us appeal to the metrics (\ref{penl}) and (\ref{penl2}). They are
already written in (4+1)-form and obeys the separating condition of the
dynamical degrees of freedom. Here constant ($\Lambda_G$) enters neither
the scalar field $(\varphi=0)$ nor the 4D metric
$(g_{\mu\nu}=\eta_{\mu\nu})$. It appears in 5D shift vector $U^{A}=\{U^5
=1,\ U^{\mu}=\epsilon (\Lambda_G /\Lambda)\eta^{\mu}\}$. However these
metrics are not gauge-invariant with respect to the above transformations
and should be rejected.

If we use the more general (4+1)-split $V^5$ with a nonholonomic basis
\cite{gl1}, it will be possible to write the metric (\ref{penl2}) (for
example for $\epsilon =1$) as
\begin{equation}\label{4.met}
   ~^{(5)}ds^2 = -(\theta^5)^2 + \eta_{\mu\nu}
   \left(dx^{\mu}- U^{\mu}(\theta^5)\right)
   \left(dx^{\nu}- U^{\nu}(\theta^5)\right)\,,
\end{equation}
where $\theta^5 =d z+A_{\mu}dx^{\mu}$, $\ U^{\mu}=\eta^{\mu}\, T_G / T$
and $A_{\mu}=f_{,\mu}$. From this extended standpoint the metric
(\ref{4.met}) satisfies the above conditions and can be considered as one
of the models in 5D General Relativity. The scalar, electromagnetic and
gravitational fields are absent here, but instead there is the shift
vector field $U^{\mu}$. The physical space-time is represented by set of
flat hypersurfaces $V^4$ which are embedded in the curved space $V^5$ with
non-vanishing exterior curvature.

\section{Conclusion}

In the paper it is shown that for the spaces of 5D General Relativity,
which admit the most symmetric 3D subspaces, the cylinder condition is
implemented dynamically. Moreover there is a thing which can be termed as
a ``spontaneous compactification'' of the fifth dimension. The regularity
condition of the metric (\ref{2.14b}) leads to a closure of $V^5 $ by the
fifth coordinate $Z$ with the period $L=2\pi\sqrt{K_{0}G} $ where
$K_{0}G>0 $. We use the parametrization of 5D metric such as (\ref{int})
which is used in the conformally invariant theory of interacting the
scalar and gravitational fields (\ref{4.4}). We suppose that such
reduction of this solution to 4D form is most natural and adequate to real
physics. The conformally invariant representation (\ref{v3.8}) is possible
for $K_{0}=1$, which gives $L=2\pi\sqrt{G}$ and $G>0$.

Thus radius $R_{G}=\sqrt{G}=L/2\pi$ of the event horizon of 5D black hole
with the metric (\ref{2.14b}), plays a role of the fundamental scale of
the theory. After the conformal transformation $\Lambda=U(1+ G/4U^2)$ the
metric takes the form (\ref{v3.8}) with $\epsilon =-1$. The system has one
classical ghost degree of freedom which is associated to scalar field
(\ref{phy1}) on the background of the flat physical metric
$\eta_{\mu\nu}$.

The Kaluza-Klein theory can be understood in some sense as a limit case of
the spaces of 5D General Relativity admitting the most symmetric 3D
subspaces. We consider the obtained solution as a ground nontrivial state
of 5D geometry. We can treat its asymmetric perturbations generated by the
classical or quantum excitations as an induced matter \cite{wes}. However
before proceeding further, it is necessary to study a stability $V^5$ with
respect to small perturbations like $\delta g_{AB}(x^{\mu})\exp{nx^5}$ by
analogy with Schwarzschild metric.

\section*{Acknowledgements}
I would like to acknowledge  Yu.Vladimirov and  M.Korkina for helpful
discussions of problems, touched in this paper.

\appendix
\section{On the metrics of constant curvature spaces}

Let us consider the flat space $E^{n+1}$ with metric:
\begin{equation}\label{a2met}
^{(n+1)}ds^2 =\epsilon_{\alpha \beta} dy^{\alpha} dy^{\beta} \quad
        (\alpha,\, \beta\,=\,1,\ldots\,,\, n+1 )\,,
\end{equation}
where $\epsilon_{\alpha \beta}= c_{\alpha}\delta_{\alpha \beta}$ (there is
no summation here!). According to \cite{ei} the basic hypersurfaces of the
second order
\begin{equation}\label{a2hyper}
  \epsilon_{\alpha \beta} y^{\alpha} y^{\beta}= e\Lambda^2
\end{equation}
were $e=\pm 1$, represent the single hypersurfaces $S^n $ of the constant
curvature $K=e/\Lambda^2 $ of the space $E^{n+1}$ with the metric
(\ref{a2met}).

The family of hypersurfaces (\ref{a2hyper}) (where a parameter of the
family is $\Lambda $) induces $(n+1)$-decomposition of $E^{n+1}$ and of
all objects on it \cite{gl1}. Let us introduce the field of the unit
vectors
\begin{equation}\label{a2-3}
 n^{\alpha}=\frac{y^{\alpha}}{\Lambda}\, , \qquad
 \epsilon_{\alpha \beta} n^{\alpha} n^{\beta}= e\,.
\end{equation}
Then
\begin{equation}\label{a2split}
   ^{(n+1)}ds^2 =  ed\Lambda^2+ ^{(n)}ds^2 \,, \qquad
      \epsilon_{\alpha \beta} = e\, n_{\alpha} n_{\beta}+ h_{\alpha \beta}
\end{equation}
where
\begin{equation}\label{a2-4}
^{(n)}ds^2=h_{\alpha \beta}dy^{\alpha} dy^{\beta}\, , \qquad
 d\Lambda=e\, n_{\alpha} dy^{\alpha}\,.
\end{equation}
Here $^{(n)}ds^2$ is the metric on the hypersurfaces
$\Lambda=\mbox{const}$ which are the hypersurfaces of the constant
curvature $K=e/\Lambda^2$.

Now we introduce the standard normalized hypersurface $S^{n}_{0}$ with the
coordinates
\begin{equation}\label{a2-5}
z^{\alpha}= n^{\alpha}= \frac{y^{\alpha}}{\Lambda}\, , \qquad
 \epsilon_{\alpha \beta} z^{\alpha}z^{\beta}= e.
\end{equation}
Then
\begin{equation}\label{a2metr}
   ^{(n+1)}ds^2 = ed\Lambda^2 + \Lambda^2 d\Omega^2\,,
\end{equation}
where
\begin{equation}\label{a2-8}
    d\Omega^2 = \epsilon_{\alpha \beta} dz^{\alpha} dz^{\beta}
  \qquad (\epsilon_{\alpha \beta} z^{\alpha}z^{\beta}= e)\,,
\end{equation}
is the metric for the space of fixed curvature $K_0=e $. Let us eliminate
from here the coordinate $z^{n+1}$. We have
\begin{equation}\label{a2-9}
  z^{n+1}=\sqrt{c_{n+1}(K_0 - S^{2}_{z})}\,, \qquad
  S^{2}_{z}\equiv \epsilon_{ik} z^{i} z^{k}\quad
        (i,\, k\,=\,1,\ldots\,,\, n )\,,
\end{equation}
\begin{equation}\label{a2-10}
    d\Omega^2 = g_{ik} dz^{i} dz^{k} \,,
\end{equation}
where
\begin{equation}\label{a2-11}
   g_{ik} = \epsilon_{ik} + \frac{c_{n+1}z^{i} z^{k}}{(z^{n+1})^2}
\end{equation}
is the metric for the space $S^n_0 $ of curvature $K_0=e $, and
$\epsilon_{ik}= c_{i}\delta_{ik}$ (there is no summation here).

As a result of stereographic projection of $S^n_0$ onto $E^n$
\begin{equation}\label{a2-12}
  z^i = x^i\left(1+\frac{K_0}{4}S^{2}_{x}\right)^{-1} \qquad
   (S^{2}_{x}\equiv \epsilon_{ik} x^{i} x^{k})\,,
\end{equation}
we obtain the following new metric tensor for $S^n_0 $
\begin{equation}\label{a2-13}
   \widetilde{g}_{kl}=
   g_{ik}\frac{\partial z^i}{\partial x^k}\frac{\partial z^j}{\partial x^l}
    = \epsilon_{kl}\left(1+\frac{K_0}{4}S^{2}_{x}\right)^{-2}\,.
\end{equation}
Finally we can write for the metric of the constant curvature space
$K=e/\Lambda^2$
\begin{equation}\label{a2-14}
 ^{(n)}ds^2 = \, ^{(n+1)}ds^2 - ed\Lambda^2
 =(\epsilon_{\alpha \beta} - e\, n^{\alpha} n^{\beta})
 dy^{\alpha} dy^{\beta} = \Lambda^2 d\Omega^2\,,
\end{equation}
where
\begin{equation}\label{a2-15}
  d\Omega^2 = \frac{\epsilon_{kl}dx^k dx^l}
  {\left(1+\frac{K_0}{4}S^{2}_{x}\right)^{2}}
\end{equation}
is the metric of space $S^n_0 $ of the unit curvature $K_0=e=\pm 1$ and
$\epsilon_{\alpha \beta} n^{\alpha} n^{\beta}= e$. In addition we note
\begin{equation}\label{a2-16}
  y^i = \frac{\Lambda x^i}{1+\frac{K_0}{4}S^{2}_{x}}\,, \qquad
  y^{n+1}= \Lambda\sqrt{c_{n+1}K_0}\
  \frac{1-\frac{K_0}{4}S^{2}_{x}}{1+\frac{K_0}{4}S^{2}_{x}} \qquad
   \left(S^{2}_{x}=\epsilon_{ik} x^{i} x^{k}\right)\,.
\end{equation}
Hence it can be seen that the above stereographic projection $S^n_0$ is
possible when $c_{n+1}=K_0$.

\def\CMPh {Commun. Math. Phys. }
\def\JPh {J. Phys. }
\def\CJP {Czech. J. Phys. }
\def\LMPh {Lett. Math. Phys. }
\def\NPh  {Nucl. Phys. }
\def\PhE  {Phys. Essays }
\def\PhL  {Phys. Lett. }
\def\PhR  {Phys. Rev. }
\def\PRD  {Phys. Rev. D }
\def\PhRL {Phys. Rev. Lett. }
\def\PhRp {Phys. Rep. }
\def\NCim {Nuovo Cimento }
\def\NuPB {Nucl. Phys. }
\def\GRG {Gen. Rel. $\&$ Grav. }
\def\CQG {Class. Quantum Grav. }
\def\prp {report }
\def\Prp {Report }
\def\GrC {Grav. $\&$ Cosmol. }
\def\DANS {Dokl. Akad. Nauk SSSR }
\def\APh {Ann. Phys. }
\def\JMM {Journ. Math. and Mech. }
\def\JMP {J. Math. Phys. }
\def\IVUZ {Izv. Vyssh. Uchebn. Zaved. Fiz }
\def\APP {Acta Phys. Pol. B }
\def\AJ {Astrph. J. }
\def\CRASP { C. R. Acad. Sci. (Paris) }


\begin{thebibliography}{99}
\bibitem{Kram1}
D. Kramer, H. Stephani, M.Maccallum and E. Herlt, Exact Solution of the
Einstein Field Equations (Energoizdat,  Moskow, 1982).
\bibitem{Bron}
K.A. Bronikov and V.N Melnikov,  \GRG  27 (1995) 465.
\bibitem{Kram2}
D. Kramer, \APP  2 (1970) 807.
\bibitem{Krech} S.S. Kokarev and V.G.
Krechet, \GrC  2 (1996) 107.
\bibitem{wes}
J.M. Overduin, P.S. Wesson, \PhRp 283 (1997) 303.
\bibitem{ei}
L.P. Eisenhart,  Riemannian Geometry (Princeton University, Princeton,
1926).
\bibitem{her}
W.C. Hernandes and G.C. Misner, \AJ (1968) 143 452;\\ M.E. Cahill and G.C.
McVittie, \JMP 11 (1970) 1382.
\bibitem{tang} F.R. Tangerlini, \NCim 27
(1963) 636.
\bibitem{pain}
P. Painlev\'{e}, \CRASP 173 (1921) 677.
\bibitem{chod}
A. Chodos and S. Detweiler, \PRD 21 (1980) 2167.
\bibitem{gl1} V.D.
Gladush, \APP  30 (1999) 3;\\ V.D. Gladush, R.A. Konoplya, \JMP 40 (1999)
955.
\bibitem{duff}
M.J. Duff, Kaluza-Klein Theory in Perspective, hep-th/9410046, 1994.
\bibitem{gl2}
V.D.Gladush, \IVUZ 11 (1979) 58 \\ (English translation:  Sov. Phys. J. 22
(1979) 1172).
\end{thebibliography}
\end{document}